\documentstyle[preprint,prb,aps]{revtex}
\begin{document}
\draft
\title{Disorder--Induced Broadening of the Density of States for 2D Electrons with Strong Spin--Orbit Coupling }

\author{A. G. Galstyan and M.~E.~Raikh}
\address{Department of Physics, University of Utah, 
Salt Lake City, Utah  84112}

\maketitle
\begin{abstract}
We study theoretically the disorder--induced smearing of the density of states in a two--dimensional electron system taking into account a spin--orbit term in the Hamiltonian of a free electron. We show that the characteristic energy scale for the smearing increases with increasing the spin--orbit coupling. We also demonstrate that in the limit of a strong spin--orbit coupling the diagrams with self--intersections give a parametrically small contribution to the self--energy. As a result, the coherent potential approximation becomes asymptotically exact in this limit. The tail of the density of states has the energy scale which is much smaller than the magnitude of the smearing. We find the shape of the tail using the instanton approach.
\end{abstract}

\narrowtext

It is well known how the random potential smears the band--edge in a 2D system. In the case of a white noise potential with a correlator
\begin{equation}
\label{cor}
<V({\bf r}) V({\bf{r^{\prime}}})> = \gamma \delta(\bf{r}- \bf{r^{\prime}})
\end{equation}
the characteristic energy scale for the smearing is
\begin{equation}
\label{eq2}
E_{2D}= \gamma \frac{m}{\hbar^{2}} \ .
\end{equation}
Deep in the tail $(E < 0 \ , |E| \gg E_{2D})$ the density of states (DOS)  falls off exponentially 
\begin{equation}
\label{2ddos}
\rho (E) \propto exp\biggl( -\xi \frac{|E|}{E_{2D}}\biggr)\ ,
\end{equation}
 where the numerical factor $\xi$ is approximately $\xi$ $\approx$ $5.8$
\cite{tho78,bre80} . The form of the tail $(3)$ follows from the instanton approach developed in Refs. \onlinecite{hal66,lan66} (see also the books \onlinecite{shk84,lif88}). The prefactor in Eq. (\ref{2ddos}), including the numerical coefficient~, was derived in Ref. \onlinecite{bre80}. In the intermediate region, $E \sim E_{2D}$, the exact form of the DOS is unknown. Within the coherent potential approximation it was studied in Ref. \onlinecite{tho78}. The autors of Ref. \onlinecite{tho78} have also performed the approximate matching of the coherent potential result and the tail (\ref{2ddos}).

Spin--orbit (SO) interaction modifies the energy spectrum of 2D electrons. The origin of this modification is either the absence of inversion symmetry in the bulk \cite{dre55,dre92} or the assymetry of the confinement potential. In the latter case the SO interaction can be taken into account by adding the following term to the Hamiltonian of a free electron \cite{ras84},
\begin{equation}
\label{hso}
\hat H_{SO}= \alpha ( \bbox{\hat \sigma} \times \bbox{k}) \bbox{n} ,
\end{equation}
where the components of $\hat {\bbox {\sigma}}$ are the Pauli matrices, ${\bf n} ||{\bf z}$ is the normal to the 2D plane, $\alpha$ is the SO coupling constant, and ${\bf k}$ stands for the electron wave vector. The energy spectrum of the Hamiltonian
\begin{eqnarray}\label{matr}
\label{ham}
 \hat H = \frac{\hbar ^{2}}{2m} k^2  +\hat H_{SO}
= \left(\begin{array}{cc}
\frac{\hbar^{2}}{2m} k^2 & \alpha ( k_{x} + i k_{y}) \\
 \alpha ( k_{x}- ik_{y}) & \frac{\hbar^{2}}{2m} k^2
\end{array}\right)
\end{eqnarray}
 consists of two branches
\begin{equation}
E_{1}({\bf k}) = \frac{\hbar^{2}k^{2}}{2m} - \alpha |{\bf k}| \ , \
E_{2}({\bf k}) = \frac{\hbar^{2}k^{2}}{2m} + \alpha |{\bf k}| .
\label{spectrum}
\end{equation}
The coressponding eigenstates have the form

\begin{equation}
\label{est}
\Psi^{(1,2)} _{{ \bf k}}({\bf r})= e^{i {\bf kr}} \chi^{(1,2)} _{\bf k} \ ,
\end{equation} 
where the spinors $ \chi^{(1,2)} _{\bf k}$ are defined as
\begin{equation}
\label{spinor}
\chi^{(1)} _{{\bf k}}= \frac{1}{\sqrt{2}} \left( \begin{array}{c} e^{ i \phi_ {{\bf k}}}  \\ -1 
\end{array}\right) \ , \  \chi^{(2)} _{{\bf{k}}}= \frac{1}{\sqrt{2}} \left( \begin{array}{c} 1 \\   e^{- i \phi_ {{\bf k}}}
\end{array}\right) .
\end{equation} 
Here $\phi_{{\bf k}}$ is the azimutal angle of the wave vector ${\bf k}$. The lower branch, $E_{1}({\bf k})$, has a minimum at
\begin{equation}
\label{k0}
k= k_{0}=\frac{\alpha m}{\hbar^{2}} \ ,
\end{equation}
with a depth
\begin{equation}
\label{depth}
\Delta=\frac{m \alpha^{2}}{2 \hbar^{2}} \ .
\end{equation} 
In the absence of a disorder the densities of states corresponding to each branch have  the form
\begin{equation}
\label{cldos}
\rho ^{(0)} _{1}(E)=\frac{m}{2 \pi \hbar^{2}} \frac{\sqrt{1+E/ \Delta} + 1 }
{\sqrt{1+E/ \Delta}} \ , \ 
\rho ^{(0)} _{2}(E)=\frac{m}{2 \pi \hbar^{2}} \frac{\sqrt{1+E/ \Delta}-1 }
{\sqrt{1+E/ \Delta}} \ .
\end{equation}
It is seen that $\rho ^{(0)} _{1}(E)$ is 1D--like, in the sense, that it diverges as $(-|E|+ \Delta )^{-1/2}$. The energy spectrum (\ref{spectrum}) and the densities of states (\ref{cldos}) are shown in Fig. 1.

The relation between the disorder and the SO coupling is measured by a dimensionless parameter 
\begin{equation}
\label{kap}
\kappa = \frac{E_{2D}}{2 \Delta} = \frac {\gamma}{\alpha ^{2}} \ .
\end{equation}
It is clear,  that if $ \kappa \gg 1$, then the spin--orbit term has a 
negligible effect on the DOS. In other words, in the limit of weak SO coupling the smearing is still determined by the energy scale $E_{2D}$. In the present paper we study the opposite 
limit of a strong SO coupling (or weak disorder), $\kappa \ll 1 $. Remarkably, in this case the DOS can be found $\em exactly$.

Let us first determine the characteristic energy scale, $E_{1D}$,  for disorder--induced broadening. Using the golden rule, the relaxation time for an electron with energy close to $E=- \Delta$ can be written as
\begin{equation}
\frac{\hbar}{\tau _{E}}\sim \gamma \rho ^{(0)}_{1}(E) \ .
\end{equation}
Then $E_{1D}$ can be found from the condition $E_{1D} \sim \hbar / \tau_{E_{1D}}$ ,  yielding
\begin{equation}
\label{e2}
E_{1D} = \frac{m}{\hbar ^{2}} (\gamma \alpha)^{2/3} \ .
\end{equation}
We see that for $\kappa \ll 1$ the new energy scale is much bigger than $E_{2D}$ but much smaller than the depth of the minimum: 
\begin{equation}
\label{uneq}
E_{1D} = \frac{E_{2D}}{\kappa^{2/3}} = \kappa ^{1/3} \Delta \ .
\end{equation} 
This last condition allows a strong simplification in the calculation of the DOS. Indeed, Eq. (\ref{uneq}) suggests  that the states in the region of smearing are composed of plane waves with magnitudes of wave vectors close to $k_0$ ,
\begin{equation}
\label{cond2} 
|{\bf k}| - k_{0} \sim \sqrt{2mE_{1D}/ \hbar^2 } \sim \kappa k_{0} \ll k_{0} \ .
\end{equation}
If we rewrite the energy spectrum $E_{1}({\bf k})$ as
\begin{equation}
\label{expansion}
E_{1}({\bf k})= -\Delta + \frac{\hbar^{2}}{2m}(|{\bf k}|-k_{0})^{2} \ ,
\end{equation} 
then Eq. (\ref{cond2}) allows to consider the second term as a small correction. The crucial observation, which allows the calculation of the DOS,
\begin{equation}
\label{dos}
\rho (E) = \frac{1}{\pi }Im \sum _{k} \frac{|\chi^{(1)} _{\bf k}|^2}{E - E_{1}({\bf k})- \Sigma_{{\bf 
k}}(E)} \ , \end{equation} 
in the closed form, is that under the condition $\kappa \ll 1$ the contribution of the diagrams with self--intersections to the self--energy, $ \Sigma _{{\bf k}}(E)$, is much smaller than the contribution of diagrams without self--intersections. In other words, in the strong SO coupling limit the  coherent potential approximation $\em becomes$ $\em asymptotically$ $\em exact$. To illustrate this statement, consider two second--order diagrams for the self--energy shown in Fig. 2. The contribution of the diagram (a) without self--intersection  to $Im \Sigma $ can be written as
\begin{equation}
\label{sig1}
Im \Sigma ^{(1)} =
\gamma ^2 Im \int \frac {d^{2}{\bf k} _{1}}{(2 \pi)^2} \int \frac{d^{2}{\bf k} _{2}}{(2 \pi)^2 } \!\ 
\frac{ |(\chi^{*(1)}_{ {\bf k}} \chi^{(1)} _{ {\bf k}_{1}})
(\chi^{*(1)}_{ {\bf k}_{1}}   
\chi^{(1)} _{{\bf k}_{2}}) | ^2}
{
\biggl (E-E_1({\bf k}_1) \biggr )^2 
\biggl(E-E_1({\bf k}_2) \biggr)} .
\end{equation} 
The contribution of the diagram (b) with self--intersection is correspondingly  
\begin{equation}
\label{sig2}
Im \Sigma ^{(2)} =
\gamma ^2 Im \int \frac {d^{2}{\bf k} _{1}}{(2 \pi)^2} \int \frac{d^{2}{\bf k} _{2}}{(2 \pi)^2 } \!\ 
\frac{ (\chi^{*(1)}_{ {\bf k}} \chi^{(1)} _{ {\bf k}_{1}})
(\chi^{*(1)}_{ {\bf k}_{1}}   
\chi^{(1)} _{{\bf k}_{1}+{\bf k}_{2}-{\bf k}_{2}})
(\chi^{*(1)}_{{\bf k}_{1}+{\bf k}_{2}-{\bf k}} \chi^{(1)} _{{\bf k}_2})  
(\chi^{*(1)}_{{\bf k}_{2}} \chi^{(1)} _{{\bf k}}) }
{
\biggl (E-E_1({\bf k}_1) \biggr )
\biggl(E-E_1({\bf k}_2) \biggr)
\biggl(E-E_1({\bf k}_1+{\bf k}_2- {\bf k}) \biggr)
} .
\end{equation} 
Noting that the scalar products $(\chi^{*(1)}_{ {\bf k}} \chi^{(1)} _{ {\bf k}^{\prime}})$ are equal to
\begin{equation}
\label{scprod}
( \chi^{*(1)}_{ {\bf k}} \chi^{(1)} _{ {\bf k}^{\prime}} ) = 
\cos \biggl (\frac{\phi _{\bf k} - \phi _{{\bf k} ^{\prime}} }{2} \biggr)
e^{- i \frac{\phi _{\bf k} - \phi _{{\bf k} ^{\prime}}}{2} } ,
\end{equation}
the integration over the angles $\phi _{{\bf k}_1} $ and $\phi _{{\bf k}_2}$ in Eq. (\ref{sig1}) can be easily performed. The main contribution to the integrals over absolute values $k_1$ and $k_2$ comes from the regions $|k_1 - k_0| \ll k_0$ , $|k_2 - k_0| \ll k_0$ . Then, using (\ref{expansion}), the energy denominators can be simplified to\begin{equation}
E-E_1({\bf k}_{1,2}) = E +\Delta -\frac{\hbar ^2}{2m}(|{\bf k}_{1,2}|-k_0)^2 .
\end{equation} 
As a result we get the following estimate for $Im \Sigma ^{(1)}$
\begin{equation}
Im \Sigma ^{(1)} \sim \gamma ^2 \frac{m}{\hbar ^2} \frac{k^{2}_0 }{|E+\Delta|^2} \ .
\end{equation} 
In contrast to Eq. (\ref{sig1}), in the second diagram the condition that the magnitudes of ${\bf k}$, ${\bf k}_1$ and ${\bf k}_2$ are close to $k_0$ restricts the integration over angles. Indeed, consider the last energy denominator, $\biggl ( E+ \Delta -\frac {\hbar ^2}{2m}(|{\bf k}_1 +{\bf k}_2 -{\bf k}|-k_0)^2 \biggr)$,  in Eq. (\ref{sig2}). It is easy to see that $|{\bf k}_1 +{\bf k}_2 -{\bf k}|$ can be close to $k_0$ only in three domains:  $\imath )$ \ $|{\bf k} _{1}- {\bf k}| \ll k_{0}$ , \ $\imath \imath )$ \ $|{\bf k} _{2}- {\bf k}| \ll k_{0}$~, $\imath \imath \imath )$ \ $|{\bf k} _{1}+ {\bf k}_{2}| \ll k_{0}$~. The size of these domains is determined from the condition
\begin{equation}
 |{\bf k}_1+ {\bf k}_2- {\bf k}| - k_{0} \sim \sqrt{m|E+ \Delta|/ \hbar^{2}} .
\end{equation}
 For the case $\imath )$, for example, this condition confines the angle $\phi_{{\bf k}_1}$ within the interval 
\begin{equation}
|\phi_{{\bf k}_1} -\phi_{\bf k}| \sim  \sqrt{m|E+\Delta|}/ (\hbar k_0) .
\end{equation}
Then the estimate for $Im \Sigma^{(2)}$ yields
\begin{equation}
\label{estimate}
Im \Sigma ^{(2)} \sim \gamma ^2 \biggl (\frac{m k_0}{\hbar ^2 |E+\Delta|} \biggr )^{3/2} .
\end{equation}
Thus, we get the following estimate for the ratio of diagrams (a) and (b)
\begin{equation}
\frac{Im \Sigma ^{(2)}}{Im \Sigma ^{(1)}} \sim  \sqrt{m|E+\Delta|}/ (\hbar k_0) \ .
\end{equation} 
In the region of broadening, $|E+\Delta| \sim E_{1D}$ , this ratio is of the order of $\kappa ^{1/3} \ll1$ . More accurate estimate (see below) gives $\kappa^{1/3} \ln (1/ \kappa)$.

Once the diagrams with self--intersections can be neglected, the summation of the remaining series is straightforward and yields the following equation for the self--energy
\begin{equation}
\label{Sigma}
Im \Sigma_{\bf k}(E) = \gamma  Im \int \frac {d^{2}{\bf k}_1}{(2 \pi)^2}
\frac{|(\chi^{*(1)}_{\bf k} \chi^{(1)}_{{\bf k}_1})|^2 }{ E - E({\bf k}_1) -\Sigma _{{\bf k}_1} (E) } \ .
\end{equation}
It is easy to see that $Im \Sigma_{\bf k}(E)$ does not depend on ${\bf k}$. Although an explicit dependence on $\phi _{\bf k}$ is present in the numerator of the integrand, it disappears after the angular integration. Substituting for $E({\bf k}_1)$ the expansion (\ref{expansion}) and performing the integration, we obtain for $Im \Sigma $
\begin{equation}
Im \Sigma = \frac{E_{1D}}{2^{4/3}} f\biggl (\frac{2^{4/3} \varepsilon}{E_{1D}} \biggr ) \ ,
\end{equation} 
where the energy $\varepsilon$ is defined as 
\begin{equation}
\varepsilon = E + \Delta - Re \Sigma \ ,
\end{equation} 
and the dimensionless function $f(x)$ satisfies the algebraic equation
\begin{equation}
f(x) = \sqrt{ \frac{x + \sqrt{f(x) ^{2}+x^{2} }}
{f(x) ^{2}+x^{2}}} \ .
\end{equation}
The function $f(x)$ is shown in Fig. 3. It turns to zero  at $x= -2^{-1/3}$. In the vicinity of  $x=- 2^{-1/3}$ it exhibits a  square--root behaviour
\begin{equation}
f (x) \simeq \frac{2^{5/6} 3^{1/2}}{5^{1/2}} \sqrt{x + \frac{1}{2^{1/3}}} \ .
\end{equation}
This behavior is usual for the coherent potential approximation.
Using (\ref{Sigma}), the density of states (\ref{dos}) can be expressed through the function $f(x)$ as follows
\begin{equation}
\rho (\varepsilon) = \frac{1}{\pi \gamma} Im \Sigma = \frac{m}{2 \pi \hbar^2} \biggl (\frac{4}{\kappa} \biggr ) ^{1/3}  f\biggl (\frac{ 2^{4/3} \varepsilon}{E_{1D}} \biggr ) \ .
\end{equation}
Clearly, the vanishing of the DOS at $\varepsilon=-2^{-5/3}E_{1D}$ is the consequence of neglecting the diagrams with self--intersections. Taking these diagrams into account leads to the smearing of this singularity and formation of the tail of the DOS. The fact that intersecting diagrams are relatively small indicates that the characteristic energy for this smearing should be much smaller than $E_{1D}$. Indeed, below we demonstrate, using the instanton approach, that the DOS in the tail has the form \begin{equation}
\label{tail}
\rho (\varepsilon) \propto exp \biggl (- \frac{\pi |\varepsilon |}{E_{2D} \ln(\Delta / |\varepsilon |)}  \biggr ) .
\end{equation} 
It is seen from (\ref{tail}) that the rate of the decay of the DOS in the tail 
is $E_{2D} \ln(\Delta / E_{2D}) \ll E_{1D}$. Note that at $|\varepsilon|\sim \Delta $, Eq. (\ref{tail}) matches the result (\ref{2ddos}) for the zero SO coupling. This conclusion could be anticipated since at energies $|\varepsilon| \gg \Delta$ the density of states does not depend on the SO coupling and Eq. (\ref{2ddos}) applies.

Within the instanton approach the density of states is given by
\begin{equation}
\label{roe}
\rho (E) \propto exp \biggl (- \frac{1}{2 \gamma}\int d^2 {\bf r}|\Phi ({\bf r})|^4 \biggr ) ,
\end{equation}
where the function $\Phi ({\bf r})$ is the solution of the nonlinear equation
\begin{equation}
\label{instantoneq}
  \hat H \Phi ({\bf r}) -| \Phi ({\bf r})|^2 \Phi ({\bf r}) =E  \Phi ({\bf r}) \ .
\end{equation} 
When the energy $E$ is close to $- \Delta$, the two--component wave function $\Phi({\bf r})$ is modulated in space with a period $2 \pi / k_0$. Then it is convenient to perform the Fourier transformation of Eq. (\ref{instantoneq}). Substituting 
\begin{equation}
\Phi ({\bf r}) =  \int d^2 {\bf r} \!\   A({\bf k}) \chi^{(1)}_{\bf k} e^{-i {\bf k}{\bf r} } \ , 
\end{equation}
we obtain
\begin{eqnarray}
\label{ak1}
&& A({\bf k})(E_{1}({\bf k})-E)= \frac{1}{(2 \pi )^{2}} \int\!\! d^2{\bf r} 
 \!\!\int \!\!d^2 {\bf k}_1\!\!\int \!\!d^2 {\bf k}_2\!\!\int \!\!d^2  {\bf k}_3 A({\bf k}_1)  A({\bf k}_2)  A({\bf k}_2) \nonumber \\ 
&&~~~~~~~~~~~~~~~~~~~~~~~~~~~~~~~~~~~~~~ \times (\chi^{*(1)}_{{\bf k}} \chi^{(1)}_{{\bf k}_1}) (\chi^{*(1)}_{{\bf k}_2} \chi^{(1)}_{{\bf k}_3}) e^{i({\bf k}_1-{\bf k}_2 +{\bf k}_3 -{\bf k} ){\bf r}} .
\label{aka}
\end{eqnarray}
Since $A({\bf k})$ depends only on the absolute value of ${\bf k}$, the angular
integration in (\ref{ak1}) can be easily performed. Using (\ref{scprod}) we obtain
\begin{eqnarray}
\label{ak2}
&& A(k)(E_{1}({\bf k})-E)= \pi^{2} \int\!\! drr 
 \!\!\int \!\!dk_1k_1\!\!\int \!\!dk_2k_2\!\!\int \!\!dk_3k_3 A(k_1)A(k_2)A(k_3) \nonumber \\
&&~~~~~~~~~~~~~~~~~~~~~~~~\times [J_{0}(kr)J_{0}(k_1r)+ J_{1}(kr)J_{1}(k_1r)]  [J_{0}(k_2r)J_{0}(k_3 r)+ J_{1}(kr)J_{1}(k_3 r)] \ ,
\label{aka1}
\end{eqnarray}
where $J_{0}(x)$ and $J_1 (x)$ are the Bessel functions of the zeroth and the 
first order respectively. Now we make use of the fact that for $E \approx -\Delta$ the magnitudes of all wave vectors in Eq. (\ref{ak2}) are close to $k_0$. More precisely, the typical range of the change of each $k$ is $|k-k_0| \sim k_{\varepsilon} = \sqrt{m(E+\Delta) / \hbar ^2}$. If we replace $k$, $k_1$, $k_2$, $k_3$ in the arguments of Bessel functions by $k_0$, then the product of Bessel functions will fall off as $r^{-2}$ and the integral over $r$ would diverge logarithmically at $r \rightarrow \infty$. This divergence should be cut at $r \sim k^{-1}_{\varepsilon}$. Then we obtain
\begin{equation}
\label{ak3}
A(k) \biggl (\frac{\hbar ^2}{2m}(k-k_0)^2 -\varepsilon \biggr)= 4k_0 \ln(k_0/k_{\varepsilon})\biggl (\int_{0}^{\infty}\!\!d k^{\prime}A(k^{\prime}) \biggr )^3 \ .
\end{equation} 
The obvious solution of this equation is
\begin{equation}
\label{solution}
A(k)= \frac{C}{\frac{\hbar ^2}{2m}(k-k_0)^2 +|\varepsilon |} \ . 
\end{equation}
Substituting (\ref{solution}) into (\ref{ak3}), we find the value of the constant $C$
\begin{equation}
\label{const}
C=\frac{1}{2 \pi^{3/2}} k^{-1/2}_0  \biggl (\frac{2m}{|\varepsilon| \hbar^2 }\biggr )^{-3/4} \ln ^{-1/2}(k_0 / k_{\varepsilon}) \ .
\end{equation}
Performing the inverse Fourier transformation, we obtain the solution of the instanton equation in the coordinate space, which is valid for $r \lesssim k_{\varepsilon} ^{-1}$:
\begin{equation}
\label{fi}
\Phi (r)= 2 \pi^2C k_0\biggl (\frac{m}{\hbar^2 |\varepsilon|} \biggr)^{1/2} \left( \begin{array}{c} J_1(k_0r) \\ -J_0(k_0r) 
\end{array}\right) \ .
\end{equation} 
Finally, upon substituting (\ref{fi}) into (\ref{roe}) we arrive at (\ref{tail}).

The above calculation allows to improve the estimate of the diagram (b) in Fig. 2. Note, that with  $A(k)$ in the form (\ref{solution}) the right--hand side of Eq. (\ref{ak1}) coincides with the expression  (\ref{sig2}) for the diagram (b). This becomes obvious if the integration over ${\bf r}$  in Eq. (\ref{ak1}) is performed explicitly. Comparing Eq. (\ref{ak3}) and Eq. (\ref{estimate}), we see that in the order--of--magnitude estimate of the diagram (b) the factor $\ln(k_0/k _{\varepsilon})$ was missing.

In conclusion, we have calculated the DOS for 2D electrons in the Gaussian random potential in the limit of a strong spin--orbit coupling. The summation of the diagram series became possible due to the fact that in the absence of disorder the energy spectrum has a minimum at some finite $k=k_0$. This causes a singularity in the bare DOS. As a result, the magnitude of smearing of the DOS increases with increasing the SO coupling. Note, that the depth of the minimum in the energy spectrum decreases in the presence of a magnetic field and disappears completely when the Zeeman splitting exceeds $2\Delta$. Then the smearing of the DOS occurs independently for both spin projections, so that the magnitude of the smearing becomes $E_{2D}$--the same as in the spinless case.

We did not analyze the real part of the self--energy, $Re \Sigma$. In fact, $Re \Sigma $  is determined by the large values of momentum and diverges logarithmically. This divergence is the consequence of the zero correlation radius of the random potential. For a small, but finite correlation radius $Re \Sigma$ causes a shift in the position of the band--edge \cite{bre80,raikh}. In the calculation of the shape of the tail this shift did not show up. This is a common situation for the instanton approach, within which the conversion from the bare to the ``physical'' energy occurs on the stage of calculating the prefactor \cite{bre80,raikh}.

The applicability of the theory developed requires the SO--induced energy scale $\Delta$ to be bigger than the inverse relaxation time $E_{2D}$ in the absence of the SO coupling. This condition seems to be met in high mobility silicon MOSFETs \cite{pud65,pud66}. According to Ref. \onlinecite{pud66}, the coupling constant $\alpha$ in this structures is $\sim 2 \cdot 10^{-6}$mev$\cdot$cm, which corresponds to $\Delta \sim 1$K. However, in the experimentally interesting situation where the metal--insulator transition occurs (see, e.g., recent references \onlinecite{krav96,pop97,sim97}), the Fermi energy lies much higher than $\Delta$.

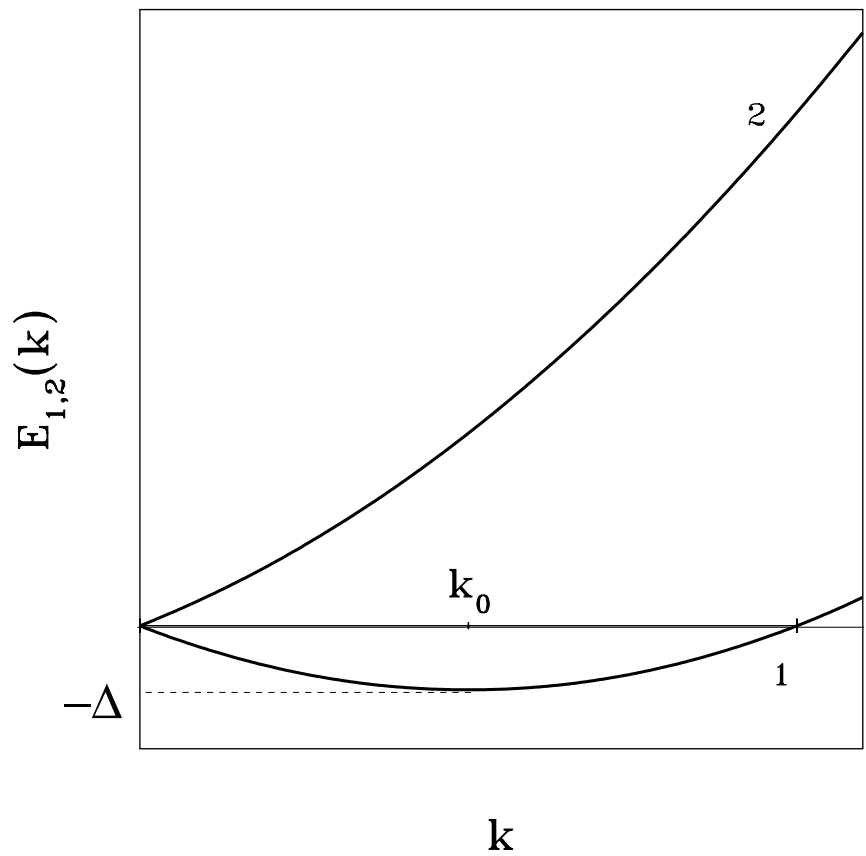
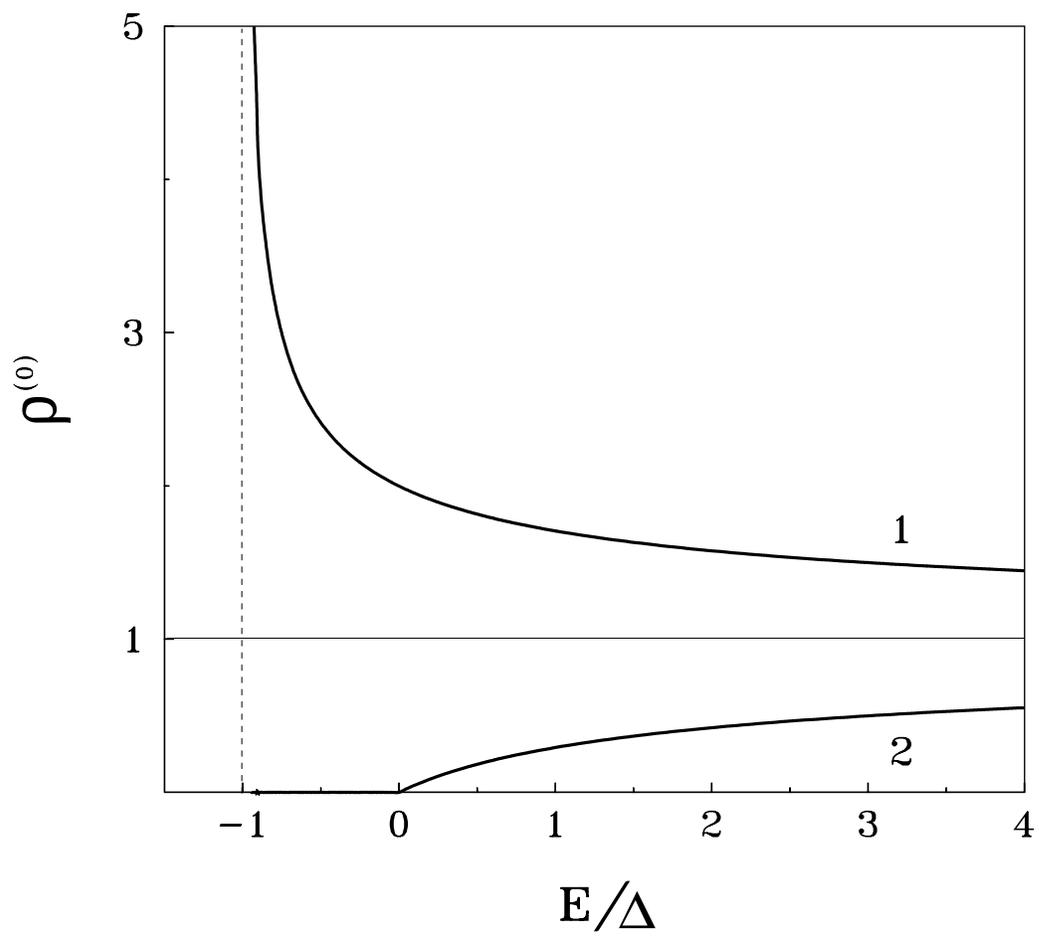
\begin{figure}
\caption{(a) The energy spectrum of a 2D system with spin--orbit coupling. (b) The  density of states  for two branches of the spectrum.}
\end {figure} 
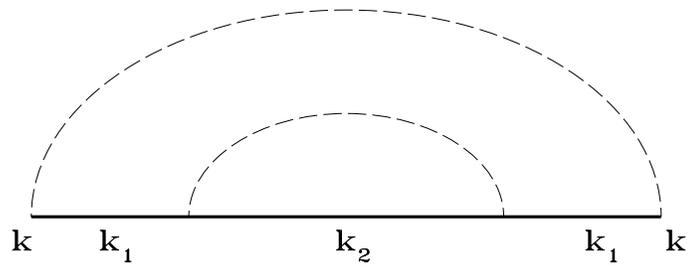
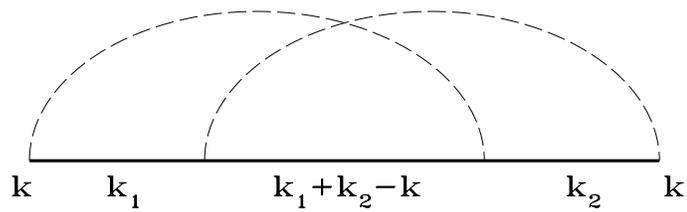
\begin{figure}
\caption{Two second--order diagrams for the self energy $\Sigma$. (a) The diagram without self--intersection. (b) The diagram with self--intersection. }
\end{figure} 
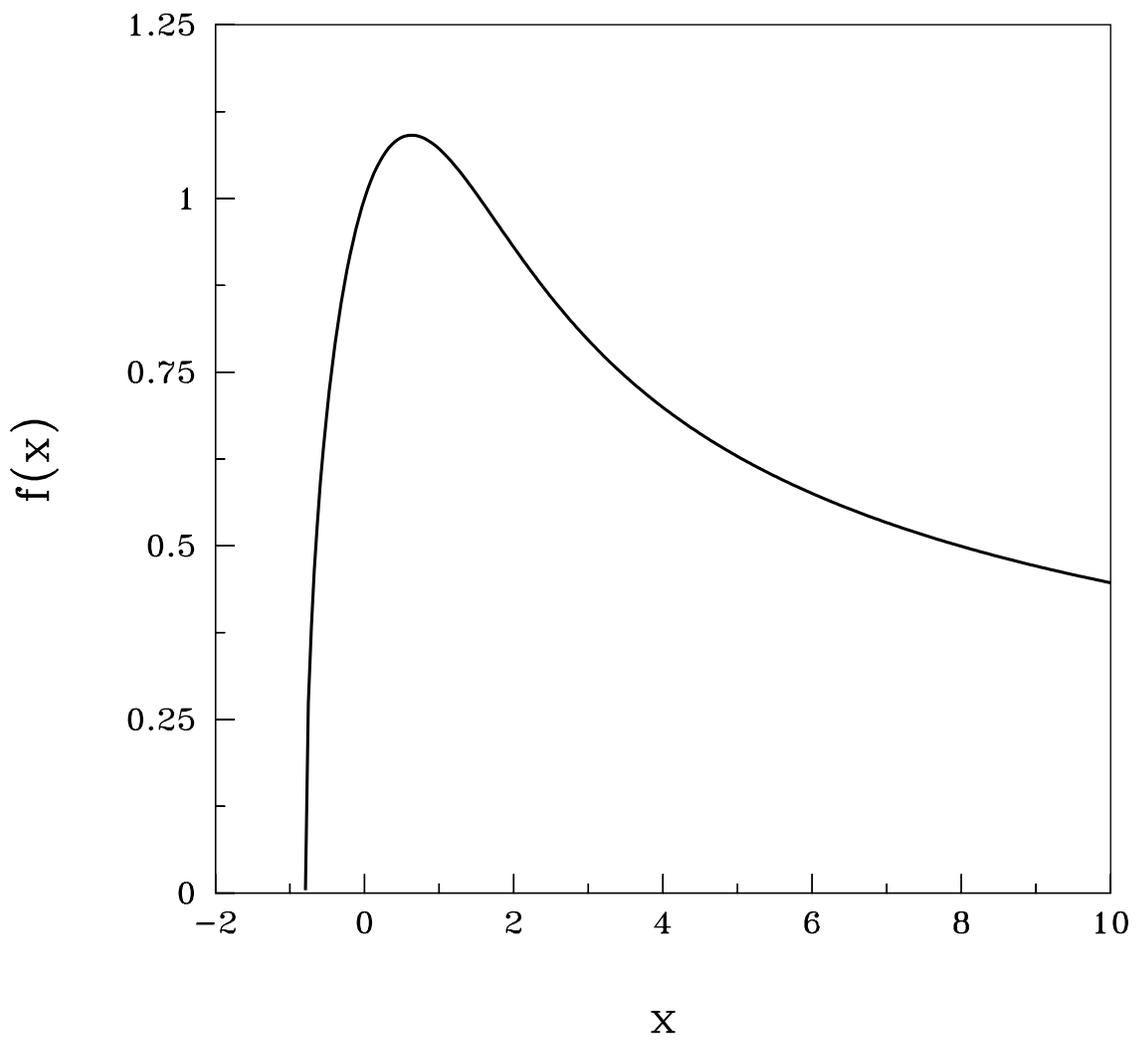
\begin{figure}
\caption{Dimensionless function $f(x)$ defined by Eq. (31). ~~~~~~~~~~~~~~~~~~~~~~~~~~~~~~~~~~~~~~~~~ }
\end{figure}
\end{document}